\documentclass[twocolumn,aps,superscriptaddress,showpacs,floatfix]{revtex4}
\usepackage{graphicx}
\usepackage{dcolumn}
\usepackage{bm}
\usepackage{CJK}

\renewcommand{\vec}[1]{\boldsymbol{#1}}

\begin{document}

\begin{CJK*}{GBK}{song}

\title{Effects of the magnetic field on the spallation reaction
implemented by BUU coupled with a phase-space coalescence afterburner}

\author{Yang Liu}
\affiliation{Institute of Modern Physics, Chinese Academy of
Sciences, Lanzhou 730000, China} \affiliation{Graduate University
of Chinese Academy of Sciences, Beijing, 100049, People's Republic
of China}
\author{Gao-Chan Yong\footnote{Corresponding author, yonggaochan@impcas.ac.cn}}
\affiliation{Institute of Modern Physics, Chinese Academy of
Sciences, Lanzhou 730000, China}
\author{Wei Zuo}
\affiliation{Institute of Modern Physics, Chinese Academy of
Sciences, Lanzhou 730000, China}

\begin{abstract}
Based on the Boltzmann-Uehling-Uhlenbeck (BUU) transport model
coupled with a phase-space coalescence afterburner, spallation
reaction of $p+^{197}Au$ at the incident beam energy of
E$_{beam}$= 800 MeV/nucleon is studied. We find that the number of
test particles per nucleon has minor effects on the neutron to
proton ratio (n/p) of the produced heavier fragments while it
affects much on their yields. The external strong magnetic field
affects the production of heavier fragments much than the n/p of
produced fragments. The n/p of free nucleons is greatly affected
by the strong magnetic field, especially for the nucleons with
lower energies.
\end{abstract}

\maketitle

%\section{Introduction}

There has been a renewed interest in the study of spallation
reactions induced by either nucleons or light charged nuclei, not
only nuclear physicists but also astrophysicists and nuclear
engineers \cite{fur00,cug07}. The spallation reaction is a kind of
nuclear reaction in which a particle (e.g. proton) interacts with
a target nucleus. Giving a high energy to the incident proton, the
nucleus is then in an excited state and can de-excite by
evaporation and/or fission. And then the high number of secondary
neutrons are produced. The fragmentation can be used as a source
of neutrons, for example, the ADS (Accelerator-Driven System)
\cite{gud99}, which is as external source to drive the
sub-critical reactor. The optimum proton energy for the production
of neutrons by spallation in a heavy metal target, in terms of
costs, target heating, and system efficiency, lies in the range
600 to 1000 MeV \cite{ads}. The spallation reaction is also
related to the Spallation Neutron Source (a new generation of
pulsed neutron sources based on the proton accelerator, and can
provide high flux pulsed neutron beams for neutron scattering.),
which is used to probe the structure of the microscopic world
\cite{sns05}.

Proton induced reactions can be described by the semiclassical
Boltzmann-Uehling-Uhlenbeck (BUU) transport model \cite{WB,LPK},
which is quite successful in characterizing dynamical evolution of
nuclear collision. However, the description of fragment formation
is not a trivial task. Because the BUU transport models are
incapable of forming dynamically realistic nuclear fragments,
certain types of afterburners, such as statistical and coalescence
models, can be used as complements. The hybrid model can be
usually used to study fragment formation, for instance, BUU+
statistical multi-fragmentation mode (SMM) \cite{arx}, BUU+
nuclear multi-fragmentation model \cite{Hkr,BAli,KHa,WPT}.
However, these models all have their own shortcomings, such as the
lack of describing the dynamical process of the fragment formation
or not giving the accurate excitation energies of fragments (which
is very important for the de-excitation of hot fragment
\cite{bot}).

In this work, we study fragment formation in proton-induced
reaction $p+^{197}Au$ at the incident beam energy of E$_{beam}$=
800 MeV/nucleon using the BUU transport model coupled with a
phase-space coalescence afterburner \cite{yong09}. We first deduce
a reasonable number of test particles per nucleon by fitting the
experimental data. Then we discuss the phenomenon of
multi-fragmentation in the external magnetic field and show how
the magnetic field affects the spallation reaction.

%\section{The theoretical model}

It is well known that the semiclassical
Boltzmann-Uehling-Uhlenbeck (BUU) transport model \cite{GF} is
quite successful in describing the evolution of nuclear collision.
The BUU equation describes time evolution of the single particle
phase space distribution function
\begin{math}f(\vec{r},\vec{p},t)\end{math}, the equation is
\begin{equation}
\frac{\partial f}{\partial t}+\vec{v} \cdot
\vec{\nabla}_{\vec{r}}f -\vec{\nabla}_{\vec{r}}\vec{U} \cdot
\vec{\nabla}_{\vec{p}}f =\bm{I}_{collision},
\end{equation}
where \begin{math}f(\vec{r},\vec{p},t)\end{math} can be viewed
semi-classically as the probability of finding a single particle
at time $t$ with the momentum \begin{math}\vec{p}\end{math} at the
position \begin{math}\vec{r}\end{math}. The right-hand side
includes the Pauli blocking. Because one has to initialize nucleon
positions to start the cascade model, nucleon positions imply a
certain density. Then we can define the mean field potential
\begin{math}U\end{math} as a function of density. In our work, we use the
Skyrme parametrization  \cite{GF}, it reads
\begin{eqnarray}
U(\rho)
=A(\frac{\rho}{\rho_{0}})+B({\frac{\rho}{\rho_{0}}})^{\sigma}.
 \label{Un}
\end{eqnarray}
Where $\sigma = 7/6$, A = -0.356 GeV is attractive and B = 0.303
GeV is repulsive. With these choices, the ground-state
compressibility coefficient of nuclear matter K= 201 MeV.
$\rho={\rho}_{n}+{\rho}_{p}$ is the baryon density and
${\rho}_{n}$, ${\rho}_{p}$ are the neutron and proton densities,
respectively.

The variations of nucleonic momentum are generally due to momentum
and spatial location dependence of its mean-field-potential $U$,
decided by the gradient force ${\nabla}_{\vec{r}} U$ in Eq.~(1).
Besides the gradient and Coulomb forces added on the charged
particles, the Lorentz force can also change the momentum of
charged particle. For the external magnetic field force of charged
nucleon felt, we employ the Lorentz force equation \cite{yongplb}.
According to Hamilton's equations, the propagations of nucleon are
\begin{eqnarray}\label{heq}
dp_{i}/dt&=&-\nabla_{r} U(r_{i})+F_{Coulomb}+F_{Lorentz},\nonumber\\
dr_{i}/dt&=&p_{i}/\sqrt{m^{2}+p_{i}^{2}}.
\end{eqnarray}

The test particle method was first applied to the nuclear Vlasov
equation by Wong \cite{wong}. In the practical BUU calculations,
to analyze fragment formation in proton induced spallation
reactions, we have used different numbers of test particles per
nucleon. From them we choose 5 and 10 test particles per nucleon
as reasonable values by comparing with experimental data. And in
the next calculations to analyze fragmentation in the reactions
with strong magnetic field, we use 8 test particles per nucleon as
the reasonable value. The number of test particles per nucleon can
not be too small as statistical fluctuations, which is inherent in
any Monte-Carlo calculation, become apparent. Also the number can
not be too large since we need to sample the phase space
distribution of nucleons \cite{GF}. Usually we use 200$\sim$ 500
test particles per nucleon to do BUU simulations, such as
nucleonic emission or meson production. Different numbers of test
particles per nucleon really less affect BUU results. However, in
the studies here, number of test particles per nucleon affects
much on the heavier fragment production. This is because small
number of test particles per nucleon gets more statistical
fluctuations back when implemented by BUU transport model. While
large number of test particles per nucleon smooths out statistical
fluctuations (which cause fragment production). In fact, the
number of produced heavier fragments is oftern very small compared
with free nucleons.

Because most BUU-type transport models are not able to form
dynamically realistic fragments, some types of afterburners, such
as statistical and coalescence models, are naturally used as a
supplement to describe nuclear multi-fragmentation. Such kind of
hybrid models can be used well to study fragment formation, for
example, in the studies of collective flows of light fragments
\cite{yong09,VK,LC,FS} and fragment formation in proton induced
reactions \cite{arx}. There are also some interesting works
available on exerting advanced coalescence models
\cite{RMa,Rsc,lwc,lwcc}. But these methods are in difficulties on
predicting heavier fragments. For the purpose of our
investigation, we take the simplest phase-space coalescence model
\cite{yong09,LC,FS}, a physical fragment is formed mainly when
nucleons with relative momenta smaller than $P_{0}$ and relative
distances smaller than $R_{0}$. The results presented in the
following discussions are obtained with $P_{0}=263$ MeV/c and
$R_{0}=3$ fm. We have changed the two parameters properly and find
that they do not affect our studied here evidently.

%\section{Results and discussions}

\begin{figure}[tbh]
\begin{center}
\includegraphics[width=0.5\textwidth]{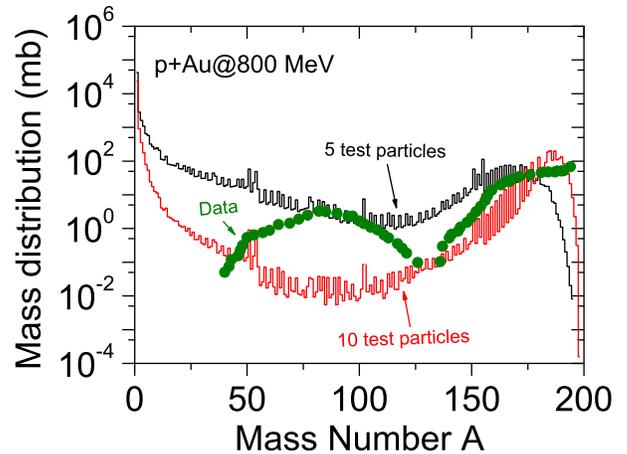}
\end{center}
\caption{(Color online) Mass distribution of fragments in the
$p+^{197}Au$ reaction at the incident energy of E$_{beam}$= 800
MeV/nucleon. Theoretical calculations (with, respectively, 5 and
10 test particles per nucleon) are compared with existing
experimental data. Data are taken from \cite{arx}.} \label{dis}
\end{figure}
For the spallation reactions, the mass distribution of fragments
are important for practical applications. Fig.~\ref{dis} shows
mass distribution of fragments in the $p+^{197}Au$ reaction at the
incident beam energy of 800 MeV/nucleon, the numbers of test
particles per nucleon used in the calculations are 5 and 10,
respectively. It is clearly shown from Fig.~\ref{dis} that both
results are roughly in agreements with the shape of the
experimental data, especially in the zone of large mass number
fragments ($A\geq130$). In the mediate range ($40<A<130$), our
results are not well in agreements with the experimental data. The
results with 5 test particles per nucleon overestimate the
production of fragments while the results with 10 test particles
per nucleon underestimate the production of fragments. From
physical point of view, small number of test particles per nucleon
gets more statistical fluctuations back when implemented by BUU
transport model \cite{GF}. Thus more fragments are produced.

As discussed above, different numbers of test particles per
nucleon really less affect the BUU results. In the studies here,
however, number of test particles per nucleon affects much on the
fragment production (especially for $20<A<150$). The small number
of test particles per nucleon gets more statistical fluctuations
back when implemented by BUU transport model \cite{GF} while large
number of test particles per nucleon smooth out statistical
fluctuations. In fact, number of produced fragments of $20<A<150$
is very small compared with free nucleons. The present studies are
the first attempts to describe the produced heavier fragments.
Further improvements of our method are in progress. Because the
experimental data are almost in the range of our simulations, we
in the next studies take 8 test particles per nucleon as our model
parameter.
\begin{figure}[tbh]
\begin{center}
\includegraphics[width=0.5\textwidth]{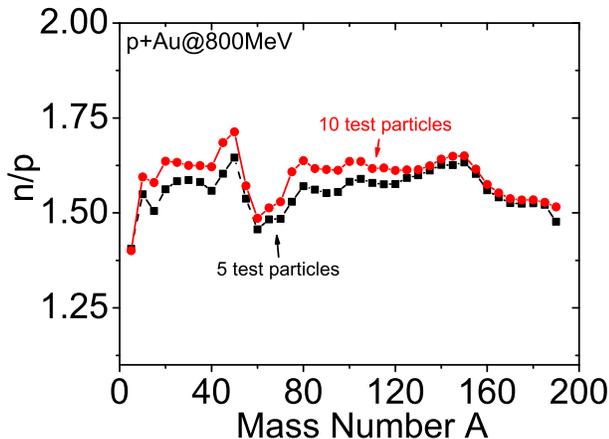}
\end{center}
\caption{(Color online) Neutron to proton ratio of nuclear
fragments with, respectively, 5 and 10 test particles per
nucleon.} \label{hdnz}
\end{figure}

Fig.~\ref{hdnz} shows neutron to proton ratio of nuclear fragments
with, respectively, 5 and 10 test particles per nucleon. We can
see that the number of test particles per nucleon has minor effect
on the neutron to proton ratio of nuclear fragments. The effects
of the number of test particles per nucleon (with 5 and 10 test
particles per nucleon, respectively) on the neutron to proton
ratio of nuclear fragments are not more than 5\%.

\begin{figure}[tbh]
\begin{center}
\includegraphics[width=0.5\textwidth]{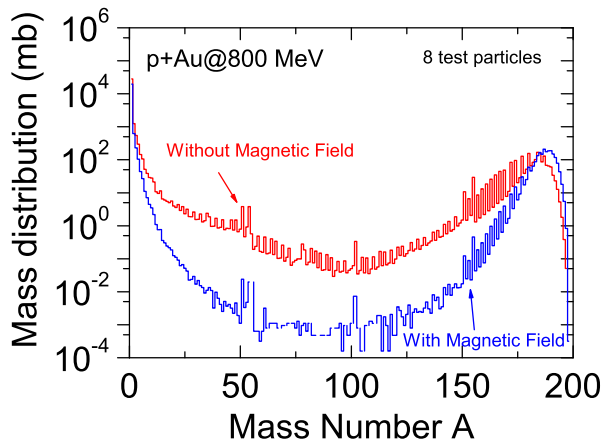}
\end{center}
\caption{(Color online) Mass distribution of fragments in the
$p+^{197}Au$ reaction at the incident energy of E$_{beam}$= 800
MeV/nucleon with and without magnetic field, the strength of
magnetic field added is 10$^{15}$ Tesla.} \label{multi}
\end{figure}
%%corrections
The condition of strong magnetic field may exist in the universe,
such as white dwarfs, neutron stars, and accretion disks around
black holes, and the maximum value of magnetic fields in the
universe may reach $10^{20}\sim 10^{42}$ G \cite{sha06}. And with
the rapid development of laser technology \cite{led03,um00},
obtaining strong magnetic field greater than 10$^{10}$ Tesla
artificially in terrestrial laboratory is possible. Also the
strong magnetic field greater than 5$\times$10$^{14}$ Tesla can be
provided via energetic heavy-ion collisions technically
\cite{car88,sk09,cassing11,cassing12}. In the following studies,
we set the strength of external magnetic field added to be
10$^{15}$ Tesla with the direction perpendicular to the reaction
plane.

To see the effects of external strong magnetic field on
multi-fragmentation, we plot Fig.~\ref{multi}, mass distribution
of fragments in the $p+^{197}Au$ reaction at the incident energy
of E$_{beam}$= 800 MeV/nucleon with and without magnetic field. We
use 8 test particles per nucleon in our simulations. From
Fig.~\ref{multi}, we can see that the magnetic field decreases
most of the formations of fragment. Fragments are formed through
statistical fluctuation in nuclear collision, the external
magnetic field prevents protons to form cluster with other
nucleons due to the Lorentz force added.
\begin{figure}[tbh]
\begin{center}
\includegraphics[width=0.5\textwidth]{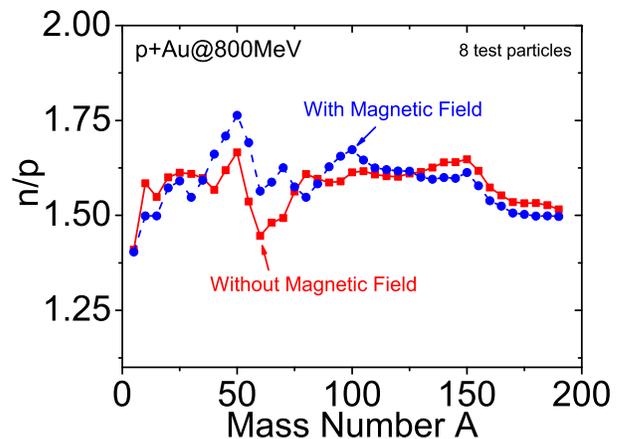}
\end{center}
\caption{(Color online) Neutron to proton ratio of nuclear
fragments in the $p+^{197}Au$ reaction at the incident energy of
E$_{beam}$= 800 MeV/nucleon with  and without strong magnetic
field.} \label{tratio}
\end{figure}
However, as shown in Fig.~\ref{tratio}, we can see that effects of
the external magnetic field on the n/p of produced fragments are
not very evident.

\begin{figure}[tbh]
\begin{center}
\includegraphics[width=0.5\textwidth]{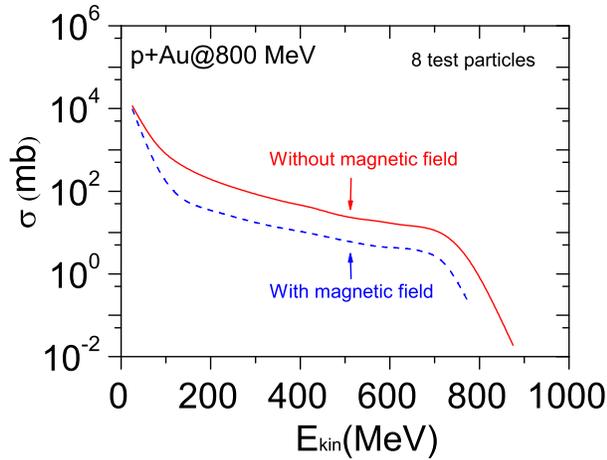}
\end{center}
\caption{(Color online) Energy spectrum of free neutron in nuclear
spallation reaction $p+^{197}Au$ at the incident energy of
E$_{beam}$= 800 MeV/nucleon with and without magnetic field.}
\label{sn}
\end{figure}
\begin{figure}[t]
\begin{center}
\includegraphics[width=0.5\textwidth]{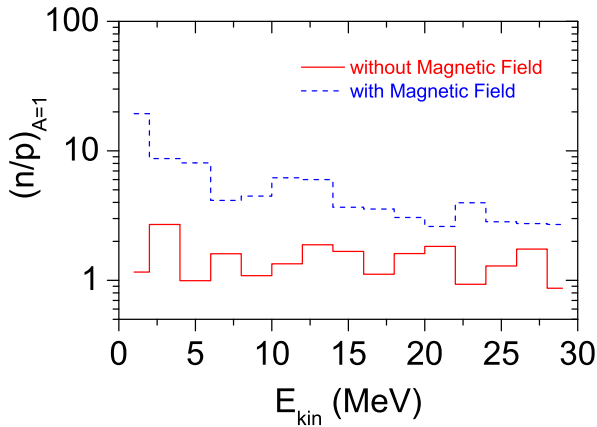}
\end{center}
\caption{(Color online) Neutron to proton ratio of free nucleons
in the $p+^{197}Au$ reaction at the incident energy of E$_{beam}$=
800 MeV/nucleon with  and without strong magnetic field.}
\label{srnp}
\end{figure}
The energy spectrum of free neutron is important for ADS and
Spallation Neutron Source. We thus studied the effects of magnetic
field on the free neutron production. Fig.~\ref{sn} shows the
energy spectrum of free neutron in nuclear spallation reaction
$p+^{197}Au$ at the incident energy of E$_{beam}$= 800 MeV/nucleon
with and without magnetic field. From Fig.~\ref{sn} we can see
that cross section of the free (especially energetic) neutrons
produced in the spallation reaction with magnetic field is smaller
than that without magnetic field. From Fig.~\ref{srnp}, we can see
that the neutron to proton ratio of free nucleons with magnetic
field is evidently larger than that without magnetic field. Free
nucleons of the spallation reaction $p+^{197}Au$ at E$_{beam}$=
800 MeV/nucleon are mainly from the initial de-excitation of hot
matter (executed by BUU through nuclear evolution). Because the
magnetic field holds protons to shoot out, the initial
de-excitation of hot matter is partly prevented. We thus see small
number of free neutrons and protons in the spallation reaction
$p+^{197}Au$ and large n/p of emitted free nucleons with external
magnetic field.

Our present coalescence method does not include fragment
de-excitation processes, the fragments are all pre-fragments
(their binding energies may not equal to their ground state
energies). These pre-fragments may have de-excitation although we
construct these fragments at the ``final stage'' of nuclear
reaction. As shown in Ref. \cite{das05}, fragment de-excitations
really affect the n/p of nuclear fragments. The effects of
fragment de-excitations on n/p ratios are evidently larger than
the effects of magnetic field on fragments' n/p ratios (shown as
Fig.~\ref{tratio} in the text). However, we can see that the
effects of magnetic field on free nucleons' n/p (shown as
Fig.~\ref{srnp} in the text) are much larger than the effects of
fragment de-excitations on n/p ratios. Since our studies here are
just the effects of the magnetic field on the fragment production
and the effects of external magnetic field on fragment
de-excitation processes are also less unknown, we thus in our
studies neglect fragment de-excitation processes. As discussed
above, the de-excitation processes of nuclear fragments do not
affect our physical results here.

As shown in Ref. \cite{cassing11}, in heavy-ion collisions there
is practically no electromagnetic field effect on observables.
This is due to the mutual dynamical compensation of transverse
components as clearly demonstrated in Ref. \cite{cassing12}. Such
compensation should be absent in a pure magnetic field and
negligible in proton-nucleus collisions. Thus the n/p ratio is
expected to be unchanged in heavy-ion collisions but its change
for proton-induced collisions in \emph{external} magnetic field
should be emerged.

%\section{Conclusions}

In conclusion, based on the transport model BUU coupled with a
phase-space coalescence afterburner, $p + Au$ spallation reaction
at the incident energy of $E_{beam}=800$ MeV/nucleon are studied.
The external strong magnetic field added may affect the formation
of fragments and the n/p of emitted free nucleons. The effects of
the Lorentz force on the n/p of heavier fragments are less
evident. These studies may be useful for nuclear physics,
astrophysics and nuclear engineering.

%\section*{Acknowledgments}

The work is supported by the National Natural Science Foundation
of China (10875151, 11175219, 10740420550), the Knowledge
Innovation Project (KJCX2-EW-N01) of Chinese Academy of Sciences,
the Major State Basic Research Developing Program of China under
No. 2007CB815004, the CAS/SAFEA International Partnership Program
for Creative Research Teams (CXTD-J2005-1).

\end{CJK*}
\end{document}